\documentclass[twocolumn,showpacs,prl]{revtex4}

\usepackage{amsmath, amsthm, amssymb}
\usepackage{graphicx}
\usepackage{dcolumn}
\usepackage{bm}

\newcommand{\tr}{\mathrm{Tr}}

\newcommand{\identity}{\openone}

\newcommand{\ie}{{\it{i.e.~}}}
\newcommand{\etal}{{\it{et al.}}}

\newcommand{\be}{\begin{equation}}
\newcommand{\ee}{\end{equation}}
\newcommand{\bea}{\begin{eqnarray}}
\newcommand{\eea}{\end{eqnarray}}

\begin{document}

\title{Thermal bound entanglement in macroscopic systems and area laws}

\author{Alessandro Ferraro$^{1}$}
\author{Daniel Cavalcanti$^{1}$}
\author{Artur Garc\'ia-Saez$^{1}$}
\author{Antonio Ac\'\i n$^{1,2}$}
\address{$^{1}$ICFO-Institut de Ciencies Fotoniques,
Mediterranean Technology Park, 08860 Castelldefels (Barcelona),
Spain\\
$^{2}$ICREA-Instituci\'o Catalana de Recerca i Estudis
Avan\c cats, Lluis Companys 23, 08010 Barcelona, Spain}

\begin{abstract}
  Does bound entanglement naturally appear in quantum many-body
  systems? We address this question by showing the existence of
  bound-entangled thermal states for harmonic oscillator systems
  consisting of an arbitrary number of particles. By explicit
  calculations of the negativity for different partitions, we find a
  range of temperatures for which no entanglement can be distilled by
  means of local operations, despite the system being globally
  entangled. We offer an interpretation of this result in terms of
  entanglement-area laws, typical of these systems.  Finally, we
  discuss generalizations of this result to other systems, including
  spin chains.
  \end{abstract}

\pacs{03.67.Mn, 03.67.-a}

\maketitle

\emph{Introduction.-}
%
Many quantum information tasks are based on several distant observers
sharing entangled pure states. However, in practical situations,
environmental noise is always present and then we unavoidably deal
with partially entangled mixed states. The first attempt to overcome
this degradation led to the idea of entanglement distillation
\cite{Benn1,Benn2}, a sequence of local operations assisted by
classical communication (LOCC) capable of extracting entangled pure
states from a large set of entangled mixed states.
%
However, in some cases, the noise degradation happens to be
irreparable: given a system composed by $n$ parties there exist
entangled states that need pure-state entanglement in order to be
generated, nonetheless no pure-state entanglement can be recovered
back by the $n$ parties via LOCC. These states are known as bound
entangled states \cite{Hor} and they represent arguably the most
striking manifestation of irreversibility in the context of quantum
information science.

Although several examples of bound entangled states have been found
\cite{HorReview}, to our knowledge they mainly have been inspired by
mathematical intuitions and simple recipes to construct them are still
lacking.  This poses doubts about whether bound entangled states,
despite their remarkable properties, are mainly a mathematical
construction.  Then up to now, a question remained open: do bound
entangled states appear inherently in nature? In particular, do these
states emerge in standard quantum many-body systems? By standard
systems we specifically mean local interacting systems in a
macroscopic thermal state characterized by only few parameters.  A
first attempt to address this question has been given in
Ref.~\cite{Geza}, where bound entanglement was detected in the thermal
state of spin systems of up to $9$ qubits.


In this work we show that bound entangled states naturally appear as
thermal states of many-body systems composed by a macroscopic number
of particles.
%
%
%
%
The fundamental intuition behind our results comes from the
entanglement-area law, a property satisfied by many condensed matter
systems according to which the entanglement between two regions scales
as the surface separating them \cite{EntMBS}. By relating bound
entanglement to area laws, we provide the first construction of bound
entangled states with a clear physical inspiration. In this sense, the
route we take goes in the reverse direction with respect to the one
usually pursued by many recent works \cite{QPT}: we borrow concepts
from the condensed matter field to get new insight on quantum
information science.

\emph{Bound entanglement and area law.-}
Consider a quantum system of $n$ particles described by a local
Hamiltonian. For the sake of simplicity,
we will restrict our analysis to translational invariant
one-dimensional systems of $n$ particles. If the system obeys an
entanglement-area law, the ground-state entanglement for a bipartite
splitting of the parties into two groups, say $A$ and $B$, scales as
the number of connections between them. This behavior has
been observed in many non-critical quantum systems, while logarithmic
corrections may appear in the critical case \cite{EntMBS}. Consider
now two different partitions of a system, one in which a contiguous
half of the particles belongs to $A$ and the other half to $B$ (we
will refer to such kind of partition as half-half), and another
partition in which the particles with even label belong to $A$ and the
others to $B$ (even-odd partition).  Because of the area law, the
entanglement will saturate for sufficiently large $n$ for the
half-half partition, while it will increase with $n$ for the even-odd
partition. In this configuration, it is reasonable to expect that, by
increasing the temperature, the entanglement in the even-odd partition
is more robust to thermal noise than in the half-half partition, and
that this behavior is preserved for large systems.

A fundamental result in the understanding of bound entanglement has
been to recognize that all distillable entangled states have a
non-positive partial transposition (NPPT) \cite{Peres}. Thus if one
finds a non-separable state with positive partial transposition (PPT)
it must be bound entangled \cite{Hor}. Now, denote by $T_{dist}^{h:h}$
($T_{dist}^{e:o}$) the threshold temperatures at which the partial
transposition with respect to all half-half (even-odd) partitions
becomes positive \cite{note1}.  Because of the area law, one can
expect that $T_{dist}^{h:h}$ is strictly smaller than
$T_{dist}^{e:o}$.  Thus, it emerges a range of temperatures for which
the system is still entangled (as detected by the entanglement in the
even-odd partition), nevertheless single particles cannot distill pure
entanglement (as the half-half partitions become PPT). Indeed, for any
pair of particles, there is always a half-half partition for which
they are in opposite sides and the partial transposition is positive
according to this splitting. In other words, bound entangled states
should appear in general under these conditions. The rest of the paper
is devoted to put on solid grounds this intuition. We first consider
systems of coupled harmonic oscillators and identify a temperature
range for which bound entanglement is present. This behavior is then
proven to persist in the macroscopic limit. We also performed the same
calculations for spin-$\frac{1}{2}$ models and found similar results.

\emph{Harmonic oscillators.-} Consider a system composed of $n$
harmonic oscillators, each one associated with position and
momentum operators $x_i$ and $p_i$ respectively ($i=1,\dots,n$),
described by the Hamiltonian
\begin{equation}\label{Hosc}
H=\frac12 \sum p_i^2+ \frac12\sum x_i V_{i,j} x_j.
\end{equation}
The diagonal elements of matrix $V$ describe the potential
energy in each oscillator, while the non-diagonal terms give the
coupling between oscillators $i$ and $j$.
In this scenario both the ground and the thermal states 
are Gaussian.  
In what follows the entanglement will be measured by the log-negativity
$E_N$, which quantifies by how much the partial transpose with respect
to a given partition fails to be positive \cite{VidWer}. Thus, when
$E_N=0$ the considered partition is PPT. In Ref.~\cite{Aud} $E_N$
between two complementary groups of oscillators, $A$ and $B$, of the
thermal state $\varrho=\exp[-H/T]/\tr\{\exp[-H/T]\}$ at temperature
$T$ was obtained:
%
%
%
\begin{equation}
E_N=\sum \log_2\{\max[1,\lambda_k(Q)]\},
\label{logneg}
\end{equation}
where $Q=P\,\omega^-\, P\omega^+$,
$\omega^\pm=W(T)^{-1}V^{\pm\frac{1}{2}}$, and $W(T)=\identity_n +
2[\exp(V^{1/2}/T) -\identity_n]^{-1}$.  We denote by
$\{\lambda_k(Q)\}_{k=0}^{n-1}$ the spectrum of the matrix $Q$
whereas $P$ is an $n\times n$ diagonal matrix with the $i$-th
entry given by $1$ or $-1$ depending on which group, $A$ or $B$,
oscillator $i$ belongs to.
%
%
An exact area law for the ground-state entanglement of this system was
proven in Ref.~\cite{Pl}. As far as for thermal states, the
entanglement for a given bipartition is upper bounded by the number of
connecting points \cite{Area}.  Here, we mainly consider harmonic
chain systems with nearest-neighbors interactions and periodic
boundary conditions.  The corresponding Hamiltonian \eqref{Hosc} is
given by a circulant potential matrix $V={\rm
  circ}(1,-c,0,\dots,0,-c)$, with $0\le c<1/2$. The system is
equivalent to a chain of harmonic oscillators coupled with a
spring-like interaction and is critical when $c\rightarrow 1/2$.

\begin{figure} {\includegraphics[width=0.5\textwidth,height=0.3\textwidth]{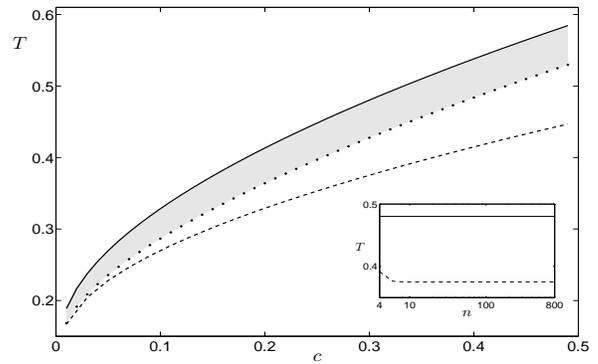}}
\caption{$T^{e:o}_{dist}$ (solid line) and $T_{dist}^{h:h}$
  (dashed line) as a function of the coupling constant $c$ for the
  Harmonic chain with nearest-neighbor interactions composed by 800
  oscillators. Notice that the range of temperatures
  $T_{dist}^{h:h}<T<T^{e:o}_{dist}$ for which bound entanglement is
  guaranteed increases when the system approaches the critical point
  $c=0.5$. The intermediate dotted line represents the upper bound to
  $T_{dist}^{h:h}$ as given by Ineq.~(\ref{Thh}) (e.g., we considered
  $m=10$ and $s=3$) and valid in the macroscopic limit: above this
  line the log-negativity in the half-half partition is zero. The
  threshold $T^{e:o}_{dist}$ in the macroscopic limit [given by
  Eq.~(\ref{negeotlim})] coincides with the one calculated numerically
  for 800 oscillators (see text for details). As a consequence, in the
  shaded region we can guarantee the presence of bound entanglement in
  the macroscopic limit. {\bf{Inset:}} $T^{e:o}_{dist}$ (solid line)
  and $T_{dist}^{h:h}$ (dashed line) as a function of the number $n$
  of oscillators composing the system (log-lin scale), for $c=0.3$
  (the same behavior is found for other values of $c$).  The gap
  $T^{e:o}_{dist}-T_{dist}^{h:h}$ is seen to remain constant with the
  size of the system, apart from an initial transient.}
\label{PhD}\end{figure}

We used Eq.~\eqref{logneg} to compute the log-negativity for the
even-odd and the half-half partition \cite{note1} for different
temperatures and number of particles. Our calculations show that the
entanglement
%
%
follows a strict area law for non-zero temperatures: it increases
linearly with $n$ for the even-odd case, while it saturates for
the half-half partition. The temperature just defines the rate the
entanglement increases with $n$ for the even-odd partition and the
entanglement saturation value for large $n$ for the half-half partition.
%
%
As shown in 
Fig.~\ref{PhD}, $T^{e:o}_{dist}$ is strictly larger than
$T^{h:h}_{dist}$, indicating, as discussed above, the presence of
bound entanglement. We performed our calculations for systems composed
by up to 800 oscillators and found that all the computed threshold
temperatures, and so also the gap $T^{e:o}_{dist}-T_{dist}^{h:h}$, are
independent of the size of the system, as can be seen in the inset of
Fig.\ref{PhD}.


\emph{Macroscopic systems.-} The results displayed in Fig.\ref{PhD}
strongly suggest that bound entanglement persists in the macroscopic
limit. Actually, we can prove this statement by establishing an
analytical formula for $E_N$ in the even-odd partition and an upper
bound for it in the half-half partition, when $n\rightarrow\infty$. We
present here the main steps of this proof. A more detailed analysis
is given in \cite{nos}.


We start with the calculation of the even-odd log-negativity,
following the one given in Ref.~\cite{Aud} for the ground state.
Consider the Hamiltonian \eqref{Hosc} with $V$ describing
nearest-neighbor interactions as shown before. Matrix $V$ is circulant
and the matrices $\omega^\pm$ can be diagonalized by a discrete
Fourier transformation, implemented by a matrix $\Omega$. Namely, we
have that
%
%
$\Omega\omega^\pm\Omega^\dagger=D^\pm$, with
$D^\pm_{k\,l}=\delta_{k\,l}\,d^\pm_k$ and
$d^\pm_k=\Lambda^{\pm1/2}_k\tanh(\sqrt{\Lambda_k}/2T)$, where
$\Lambda_k=1-2c\cos{\left(2\pi k/n\right)}$ are the eigenvalues of
$V$.
Concerning matrix $P$, one has that:
\begin{equation}
\Omega\,P\Omega^\dagger=\tilde{P}=\left(\begin{array}{cc} 0 & \identity_{n/2} \\
    \identity_{n/2} & 0\end{array}\right)\;.
\end{equation}
The spectrum of $Q$ coincides then with the one of
$\tilde{P}D^-\tilde{P}D^+$, which in turn is straightforward to
calculate due to its block diagonal structure.
%
%
Defining the function
\begin{equation}
f(k,n,c,T)=\sqrt{\frac{\Lambda_{k+n/2}}{\Lambda_{k}}}
\tanh\frac{\sqrt{\Lambda_k}}{2T}
\tanh\frac{\sqrt{\Lambda_{k+n/2}}}{2T} 
\nonumber
\end{equation}
one has that the eigenvalues of $Q$ that can contribute to the
log-negativity \eqref{logneg}
%
%
are given by $f(k,n,c,T)$, with double multiplicity
and $k=0,\dots,n/4$ (for $n$ multiple of $4$, $n\ge4$).
%
%
The log-negativity of the even-odd partition is different from
zero when the temperature $T$ is such that $f(0,n,c,T)>1$.
In particular, the curve $f(0,n,c,T)=1$ gives the threshold
temperature $T^{e:o}_{dist}$, in formula:
\be \sqrt{\frac{1+2c}{1-2c}}
\tanh\left(\frac{\sqrt{1-2c}}{2T}\right)\tanh\left(\frac{\sqrt{1+2c}}{2T}\right)=1\,.
\label{negeotlim} \ee
The threshold above coincides with the one depicted in
Fig.~\ref{PhD} (solid line) and it is independent on the total
number of particles $n$ (see also the solid line in the inset of
Fig.~\ref{PhD}), \ie it holds 
in the macroscopic limit. For temperatures below $T^{e:o}_{dist}$
there exists a $\overline{k}(n,c,T)$ such that $f(k,n,c,T)>1$ for
$k<\overline{k}(n,c,T)$, which in turn gives rise to the following
expression for the log-negativity:
\begin{equation*}
E_N=\!\!\!\!\sum_{k=0}^{\overline{k}(n,c,T)} \!\!\log_2 f(k,n,c,T)
\simeq\frac{n}{2\pi}\int_{0}^{\overline{x}(c,T)} \!\!\!\!\!\!\!\!\!\!dx\log_2f(x,c,T),
\end{equation*}
where, for large $n$, we have replaced the sum over $k$ by an integral
over $x=2\pi k/n$.
As a consequence we see that the log-negativity grows
linearly with the system size also for non-zero temperatures.

Regarding the half-half partition
%
%
it is possible to find an exact upper bound for the threshold
temperature $T^{h:h}_{dist}$. Remarkably, this allows to identify
a range of temperatures for which the presence of bound
entanglement can be guaranteed also in the macroscopic limit.
We proceed as follows. Let us define the matrix
\begin{equation}
X_{ij}=\omega^-_{ij}\sum_{k=0}^{n/2-1}\sum_{h=n/2}^{n-1}
(\delta_{ik}\delta_{jh}+\delta_{jk}\delta_{ih})\,
\end{equation}
where $\delta_{ij}$ denotes the Kronecker delta. Following
Ref.~\cite{Area}, the log-negativity 
is zero when
\begin{equation} \lambda_{\rm min}[W(T)]^{-2}+2\,{\rm
max}_i|\lambda_i[X\omega^+]|<1\;, \label{cineq}
\end{equation}
where 
\begin{equation}
\lambda_{\rm
min}[W(T)]=\frac{e^{\sqrt{1+2c}/T}+1}{e^{\sqrt{1+2c}/T}-1}.
\label{lmin}
\end{equation}
Recognizing that the second term in the left hand side of
(\ref{cineq}) is twice the spectral radius $r(X\omega^+)$ of the
matrix $X\omega^+$, we can use any matrix norm to bound it from above
\cite{HJ}. An upper bound for $r(X\omega^+)$ is then given by
$r(X\omega^+)\le||X\omega^+||\le||X||\,||\omega^+||$. We consider the
maximum row sum matrix as a representative norm: $||A||\equiv{\rm
  max}_i{\mathrm \sum_j} |A_{ij}|$ .
The goal, now, is to bound $||\omega^+||$ and $||X||$. Before
proceeding, recall that $\omega^\pm$ are circulant matrices, hence
completely specified by their first row $\omega^\pm={\rm
  circ}(v_0^\pm,\dots,v_{n-1}^\pm)$.
%
%
One can show that 
%
%
\begin{equation}
v_l^\pm=\frac{1}{2\pi}\int_0^{2\pi}dx\,d^\pm(x) e^{i x l}\;,
\label{vls}
\end{equation}
%
when $n\rightarrow\infty$.  As a consequence, for any integer $s$, by
integrating by parts $s$ times we have:
\begin{eqnarray}
|v_l^\pm|
&\le&\frac{1}{2\pi l^s} \int_0^{2\pi}dx\,
\left|\frac{d}{dx^s}d^\pm(x)\right|\equiv\frac{C_s^\pm}{2\pi l^s}\,.
\label{Cpm}
\end{eqnarray}

Let us bound first $||\omega^+||$. Being $\omega^+$ a
circulant matrix, it follows that there is no need to look for the
maximum over the rows. 
%
Then one can write, for any integer $m$, $||\omega^+||=S^+_m+{\cal
E}_m^+$, where we defined the partial sum and the residual term,
respectively, as follows:
\begin{equation}
S^{+}_m\equiv\sum_{l=-m}^{m}\!\!|v_l^{+}|\;,\qquad {\cal
E}_m^{+}\equiv2\sum_{l=m+1}^{\infty}\!|v_l^{+}|\;.
\end{equation}
In order to obtain a bound on $||\omega^+||$ one can now fix $m$,
calculate explicitly $S^+_m$ and bound ${\cal E}_m^+$ from above.
This latter step can be
achieved by using the bound in Eq.(\ref{Cpm}), leading to ${\cal
  E}_m^+ \le C_s^+\zeta(s,m+1)/\pi$
%
(where $\zeta(s,m+1)$ is the generalized Riemann zeta function).
Summarizing, for any integer $s$ and $m$, we have proven that 
\be
||\omega^+|| \le S^+_m+\frac{C_s^+}{\pi}\zeta(s,m+1)\equiv K_{m,s}^+\,,
\label{wbound} \ee 
the bound above being tighter for large $m$ and
$s$.

Concerning the upper bound for $||X||$ we have that $||X||$
coincides with $||B||$ where $B$ is the symmetric Toeplitz matrix
with the first row given by $(v_{n/2}^-, v_{n/2-1}^-, \dots,
v_2^-, v_1^-)$. We then need to determine which row of $B$ has the
maximum sum. Fortunately, when $n\rightarrow\infty$ the periodic
boundary conditions can be disregarded and the first row of $B$
determines its norm. This is because going from the first to the
second row we simply remove the term $|v_1^-|$, and so on for the
other rows.
%
Similarly as we did for $||\omega^+||$ we obtain that for any integer
$s$ and $m$
\be
||X|| \le S^-_m+\frac{C_s^-}{2\pi}\zeta(s,m+1)\equiv K_{m,s}^-\,,
\label{xbound}
\ee
where $S^-_m\equiv\sum_{l=1}^{m}|v_l^-|$. As done before, $S^-_m$ can be
calculated explicitly and the bound is tighter for large $m$ and
$s$.
Summarizing, considering Eqs.~(\ref{cineq}), (\ref{lmin}),
(\ref{wbound}), and (\ref{xbound}), we have shown that, in the
macroscopic limit, the log-negativity in the half-half partition
is zero when the following inequality is satisfied:
\begin{equation}\label{Thh}
2K_{m,s}^+K_{m,s}^-+\left(\frac{e^{\sqrt{1+2c}/T}-1}{e^{\sqrt{1+2c}/T}+1}\right)^2<1\,.
\end{equation}
Based on the formula above and on the threshold (\ref{negeotlim}), we
depicted in Fig.~\ref{PhD} the region in the $c-T$ plane for which
bound entanglement is present in the macroscopic limit (shaded
region). We see that for any coupling we can guarantee that there
is a range of temperatures for which the log-negativity in the
half-half partition is zero, nevertheless the state is entangled. Thus
the temperature, a single macroscopic and measurable quantity, clearly
determines the distillability properties of the system.
\begin{figure}
{\includegraphics[width=0.5\textwidth,height=0.3\textwidth]{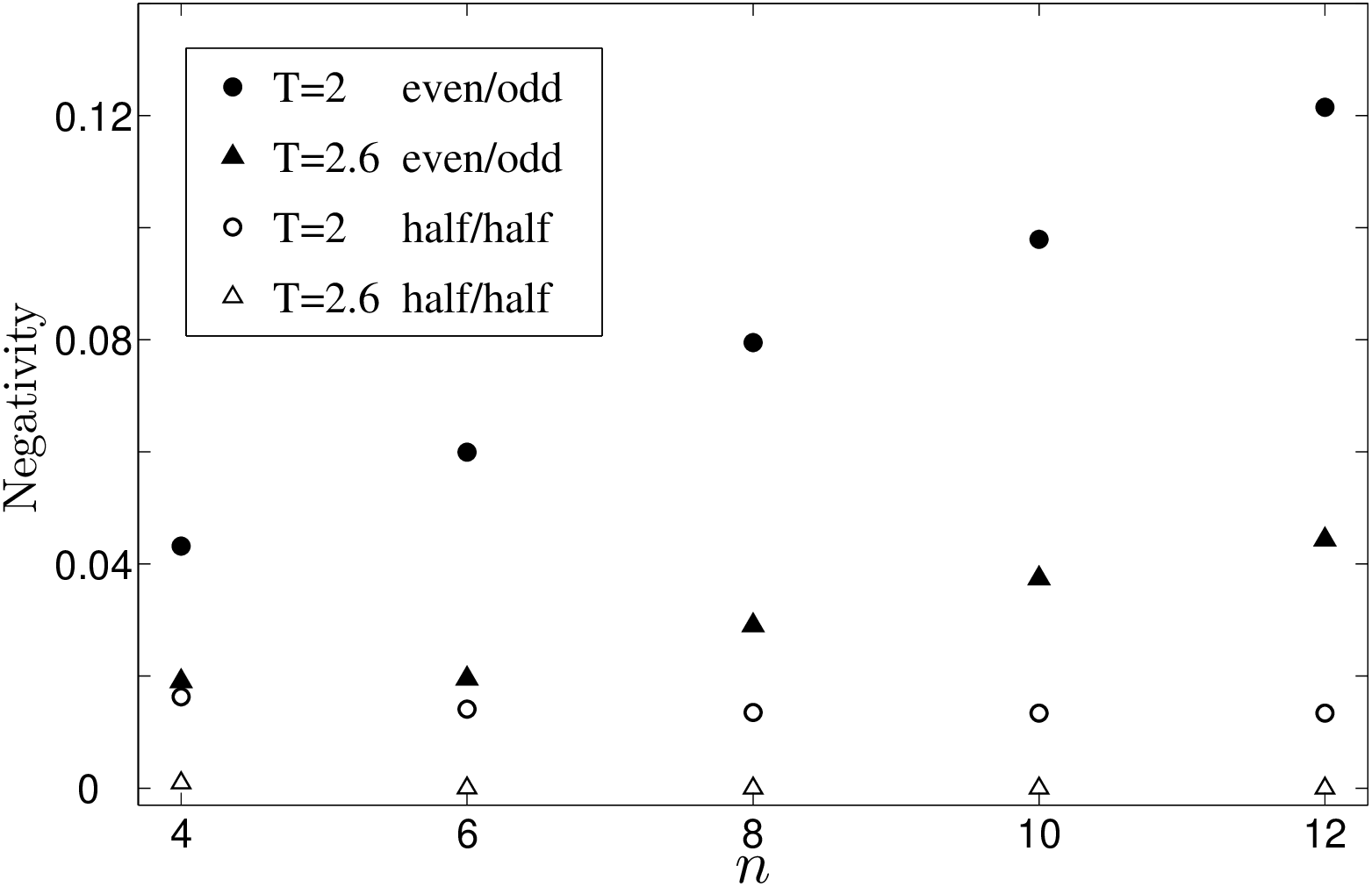}}
  \caption{Negativity in the even-odd (full symbols) and half-half
    (empty symbols) partitions for thermal states of $n$ spin-one-half
    particles with nearest-neighbor Hamiltonian given by $H_{XX}=-\sum
    (\sigma^x_{i}\sigma^x_{i+1}+\sigma^y_{i} \sigma^y_{i+1}) +B\sum
    \sigma_i^{z}$, and $B = 1.9$. The temperatures for each partition
    are set to be, from top to bottom, $T =2,2.6$ . In the even-odd partition we
    can clearly see an increase of the negativity with respect to the
    system size, whereas it saturates in the half-half partition. Both
    behaviors are expected from the entanglement-area law. For $T=2.6$
    we see that the state is bound entangled (the negativity in the
    half-half partition being zero).}
  \label{XYeo}
\end{figure}

\emph{Concluding remarks.-}
We remark that our results, when combined with those of Ref.~\cite{AW},
prove that entangled states such that all bipartitions are
PPT, cannot be obtained for the harmonic systems studied here
\cite{note2}. This is because the threshold temperatures for the
even-odd partition given above coincide with those
found in Ref.~\cite{AW} for full separability. Thus,
thermal states become PPT and fully separable at the
same temperature. 

Finally, we have considered other models of harmonic chains as well as
spin systems.  All the obtained results are consistent with the
previous reasoning \cite{nos}: there is a temperature range for which
the negativity in the half-half partitions is zero, nevertheless the
system is still entangled as proven by the even-odd negativity. For
spin systems we could not go beyond 12 particles due to computational
hardness, but the observed numerical results again support the
existence of thermal-state bound entanglement in the macroscopic limit
(see Fig.~\ref{XYeo}). Due to the generality of the area laws, the
present results are expected to hold for a variety of other systems.
These findings, besides showing novel aspects of entanglement in
many-body systems,
are of relevance from an application-oriented perspective. As shown
here, quantum correlations emerge ``gradually'' in a many-body system:
starting at high temperatures in a classically correlated scenario,
the system passes through a regime where quantum correlations start to
be present, but in a bound form. Then, for lower temperatures, the
system enters a truly quantum correlated regime where free distillable
entanglement is present and potentially available for quantum
information applications.

\emph{Note added}.- By completing this work, bound entanglement was
independently found in three-qubit reduced states of the $XY$ model at
nonzero temperature \cite{patane}. Here we have shown the existence
of bound entanglement in the whole thermal state of the macroscopic
system.

\begin{acknowledgements}
  This work is supported by the EU QAP project, the Spanish MEC, under
  FIS2004-05639 and Consolider-Ingenio QOIT projects, and a ``Juan de
  la Cierva" grant, the Generalitat de Catalunya, and the Universit\'a
  di Milano under grant ``Borse di perfezionamento all'estero".
\end{acknowledgements}


\begin{thebibliography}{30}
\bibitem{note1} In fact, due to the translational invariance of the systems considered here,
all the half-half partitions are equivalent.
\bibitem{note2} Fully PPT entanglement
has been found in spin systems up to 9 qubits in Ref.~\cite{Geza}.
\end{thebibliography}
\end{document}